\documentclass[12pt, preprint]{aastex}

\usepackage{lineno}
\usepackage{natbib}

\shorttitle{Escape from Vela~X}

\begin{document}


\title{Escape from Vela~X}

\author{
  J.~A.~Hinton\altaffilmark{1},
  S.~Funk\altaffilmark{2},
  R.~D.~Parsons\altaffilmark{3},
  S.~Ohm\altaffilmark{1,3}
}

\altaffiltext{1}{Department of Physics and Astronomy, University of
  Leicester, Leicester LE1 7RH, UK} 
\altaffiltext{2}{Kavli Institute for Particle Astrophysics and
  Cosmology, SLAC, 2575 Sand Hill 
Road, Menlo Park, CA-94025, USA, funk@slac.stanford.edu} 
\altaffiltext{3}{School of Physics \& Astronomy, University of Leeds,
  Leeds LS2 9JT, UK} 

\begin{abstract}

  Whilst the Vela pulsar and its associated nebula are often
  considered as the archetype of a system powered by a $\sim$$10^{4}$
  year old isolated neutron star, many features of the spectral energy
  distribution of this pulsar wind nebula are both puzzling and
  unusual. Here we develop a model that for the first time relates the
  main structures in the system, the extended radio nebula (ERN) and
  the X-ray cocoon through continuous injection of particles with a
  fixed spectral shape. We argue that diffusive escape of particles
  from the ERN can explain the steep Fermi-LAT spectrum. In this
  scenario Vela~X should produce a distinct feature in the
  locally-measured cosmic ray electron spectrum at very high
  energies. This prediction can be tested in the future using the
  Cherenkov Telescope Array (CTA). If particles are indeed released
  early in the evolution of PWNe and can avoid severe adiabatic
  losses, PWN provide a natural explanation for the rising positron
  fraction in the local CR spectrum.
\end{abstract}


\section{Introduction}

\begin{figure*}[htb]
\epsscale{1.}
\plotone{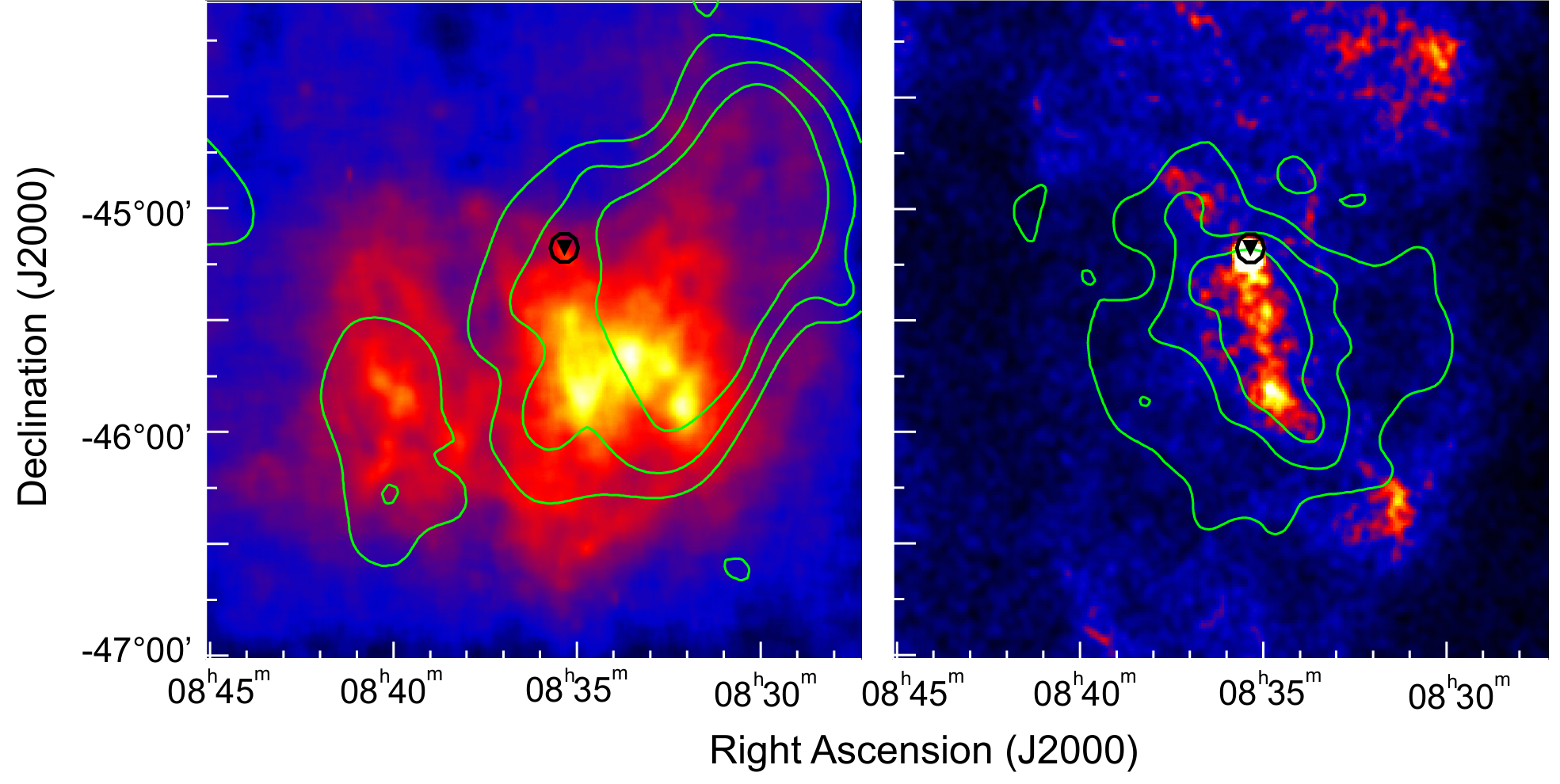}
\caption{ Left: Radio image of a $3^{\circ}\times3^{\circ}$ region
  around Vela (black triangle), from 8\,GHz 
  VLA data~\citep{Frail:1997p3296}. Overlaid are Fermi-LAT
  4,5 and 6 $\sigma$ significance contours (green) above 800
  MeV~\citep{fermi:VelaX}. Right: ROSAT (0.9--2.4~keV) count
  map~\citep{Markwardt:1997p3699}, smoothed 
  with a Gaussian kernel of 2.25', saturated at 10\% of the pulsar
  peak intensity. Overlaid are 
  H.E.S.S.\ smoothed excess contours (25, 50, 75\% of the peak
  emission)~\citep{hess:VelaX}. The black circle
  indicates the 4' region where most of the hard X-ray emission
  originates~\citep{Mangano:2005p2719}.
\label{fig:1}
}
\end{figure*}

The proximity \citep[distance: 290~pc, measured by parallax
][]{Caraveo:2001p930,Dodson2003} of the Vela pulsar (PSR\,B0833$-$45)
has allowed it to be studied in great detail across the whole
electromagnetic spectrum. PSR\,B0833$-$45 (spin-down power $\dot{E} =
7 \times 10^{36}$ erg/s, characteristic age $\tau_{c}$ = 11,000
years~\citep{atnf})
is embedded in a diffuse radio nebula~\citep{Large:1968p3223}.
Higher angular-resolution radio~\citep{Dwarakanath:1991p3369,
  Duncan:1996p3426} and X-ray~\citep{Kahn:1985p3493,
  Aschenbach:1995p669} observations of the region established the
presence of (1) a large circular shell (about 8$^{\circ}$ in
diameter) -- the Vela SNR -- and (2) an extended radio nebula (ERN) of
size $2^{\circ} \times 3^{\circ}$ in the centre -- dubbed
Vela~X (see Figure~\ref{fig:1} left). Based on the high degree of
polarisation and the flat radio spectral index
\citet{Weiler:1980p1759} suggested that Vela~X is a
\emph{pulsar wind nebula} (PWN) powered by PSR\,B0833$-$45. Such PWNe
act as calorimeter, containing the time-integrated particle outflow
and magnetic field from the pulsar.
The radio emission is far from uniform, exhibiting a network of
filaments throughout the nebula~\citep{Frail:1997p3296} . The
brightest of these non-thermal filamentary structures in Vela~X
emerges to the south of PSR\,B0833$-$45
and has a size of $45' \times 12'$ (Figure~\ref{fig:1} (right).
The X-ray counterpart to this bright radio filament was detected using
ROSAT~\citep{Markwardt:1995p3614} (see Figure~\ref{fig:1} right) and
dubbed ``the cocoon''. It
is thought to be the result of the interaction of the PWN with the SNR
reverse shock ~\citep{Blondin:2001p640}.

X-ray observations performed by
\citet{Mangano:2005p2719} found that within $1'$ of the Vela pulsar
the 3--10~keV spectrum softens with increasing distance,
a signature of electrons cooling via synchrotron emission. BeppoSax
data demonstrated that the nebula emission in the inner $12'$ in
radius extends up to $\sim$200 keV~\citep{Mangano:2005p2719} with most
of the emission coming from within $4'$ of the pulsar, suggesting
particle acceleration close to the pulsar (black circle in
Figure~\ref{fig:1}).


H.E.S.S.\ detected TeV $\gamma$-ray emission from Vela~X exhibiting an
atypical spectrum with energy flux peaking at $\sim$10
TeV~\citep{hess:VelaX}. The TeV emission is dominated by an elliptical
region of $58' \times 43'$, extending significantly beyond the X-ray
cocoon, yet smaller than the ERN (see Figure~\ref{fig:1} right). The
latest H.E.S.S.\ results~\citep{hess:velaXNew}, show significant, but
low-surface brightness TeV emission beyond this region.
Perhaps surprisingly, the $\gamma$-ray spectrum in
the outer parts is identical to that of the central region. The
H.E.S.S.\ detection of emission from the cocoon provided the first
robust estimate of the magnetic field strength in this region: $\sim 4
\mu$G. No excess TeV emission was seen from the hard X-ray
emitting region close to the pulsar,
implying a magnetic field of $B>50\mu$G in the immediate vicinity of
the pulsar, consistent with the cooling signature detected with
BeppoSax.

Finally, Fermi-LAT observations in the Vela pulsar off-pulsa data
established the presence of an extended $\gamma$-ray emitting
structure at energies above 800~MeV spatially coincident with the ERN
(see Figure~\ref{fig:1} left) with a spectral index of $\sim
2.4$~\citep{fermi:VelaX}. The two-peaked nature of the $\gamma$-ray
spectral energy distribution (SED) led the authors to
conclude~\citep[following an earlier suggestion by][]{OkkieVelaX} that
there are two distinct populations of electrons, one responsible for
the radio and GeV $\gamma$-ray emission and the other for the X-ray
and TeV emission. The magnetic field derived through these
observations is $\sim 4 \mu$G.

Despite the wealth of experimental data and several theoretical
studies, many questions remain about the relationship between the
emission seen on different scales in Vela~X. In particular, there are
several important observational facts for which no satisfactory
explanation has yet been put forward. Firstly, the steep GeV spectrum,
indicating an unexplained absence of $>100$ GeV energy particles from
the ERN. Secondly, the relative
dimness of the TeV Nebula. Comparing the spin-down luminosity of the
Vela pulsar to the TeV luminosity above 500~GeV ($10^{33}$ erg/s)
yields an efficiency in the cocoon of 0.01\%, significantly lower than
typical TeV PWNe~\citep{hess:VelaX}. In addition, the lack of spectral
variations across Vela~X above 1~TeV is not readily explained in any
model in which the TeV size is limited by cooling.

Here we show that a scenario in which the high energy particles have
diffused out of the ERN, and where the cocoon is a relatively recent
feature of the PWN, may answer these questions. This scenario has
implications for the locally measured electron spectrum at high
energies. It has been pointed out~\citep{Shen1970} that
only very local sources can significantly contribute to the flux of
cosmic ray electrons above 1~TeV, with Vela (and Geminga) long seen as
likely candidates~\citep[see
e.g.\ ][]{Nishimura1980, AtoyanAharonianVoelk1995, Kobayashi2004}. The
recent measurements of a rise in the positron fraction above $\sim
10$~GeV~\citep{pamela:positrons, fermi:positrons}
indicate that a relatively local source of positrons must be present.
Pulsars and their nebulae as plausible candidates~\citep[see
e.g.][]{ChiChen1996, Buesching2008, Profumo2008, YukselKistler2009},
given that we see high-energy electrons/positrons radiation at TeV
energies. 
However, the details of the escape of particles from the
PWN region have been left open. Here we demonstrate for the first time
how the signature of escape may be directly visible in the SED of Vela~X.


\section{Model}

We explore the possibility that particle propagation effects dominate
the spectral and morphological features of Vela~X. We attempt to
reconcile the properties of the TeV and GeV nebula by considering an
electron injection from the pulsar wind termination shock which
represents a fixed fraction of the spin-down power of the pulsar and
has a fixed spectral shape. The two $\gamma$-ray-emitting zones of
Vela~X are discussed in turn.

\paragraph{The TeV Nebula:} Whilst the TeV emission of Vela~X is now
known to extend over much of the ERN region, the bulk of the
TeV-emitting particles appear to be confined to a small fraction of
its the volume. This emission is peaked in the cocoon, but a
consistent spectral shape is measured throughout the ERN. This
suggests that the electron population of the cocoon is not
cooling-limited in extent and that the probability of escape from the
system is very small and/or energy-independent. These properties
appear consistent with the paradigm of (at least young) PWNe as
internally advection-dominated systems from which particles cannot
easily escape~\citep{AharonianAtoyanKifune1997}. The energetic
particle content should then be consistent with the power supplied by
the pulsar during the period of injection into the cocoon. As the
energetic particle content of Vela~X is well-determined by TeV and
X-ray observations, the age of the TeV cocoon can be expressed as:
$t_{\mathrm{coc}} = E_\mathrm{coc} / (\dot{E}_ {\epsilon}) \approx 70
\mathrm{yrs}/\epsilon$, with $E_{\mathrm{coc}} \sim 1.5 \times
10^{46}$ erg~\citep{fermi:VelaX}. Whilst there are considerable
uncertainties on the efficiency factor $\epsilon$ of converting
spin-down power into relativistic electrons for any plausible value of
$\epsilon$, the cocoon is much younger than the system as a
whole. This is consistent with an interpretation of the cocoon as a
feature arising from the interaction with the SNR reverse
shock~\citep{Blondin:2001p640}.
The maximum energy of the cocoon particles is constrained by the
H.E.S.S.\ measurements to $\sim$70 TeV~\citep{hess:VelaX}. This
maximum energy has been interpreted as a consequence of cooling of
higher energy particles in the cocoon~\citep{okkieBirthPeriod,
  fermi:VelaX}. However, the synchrotron lifetime for 70~TeV electrons
is $t_{\rm sync} = 10^4 (B/4\mu \mathrm{G})^{-2}$ years, consistent
with the age of the Vela pulsar but not with the lifetime of the
cocoon. We hypothesize that the maximum energy of cocoon particles is
determined by cooling in the high-$B$-field ($\sim 100 \mu$G) region
in the immediate ($\sim 0.5$ pc) vicinity of the pulsar \emph{before}
injection into the low-$B$-field cocoon region. The $O(10\mathrm{yr})$
residence time required for a cooling break to appear at 70~TeV for
particles emerging from the high B-field region is consistent with
expectations for the post-shock flow~\citep[see
e.g.][]{Mangano:2005p2719}.

\paragraph{The Extended Radio Nebula} It seems likely that the ERN
contains particles injected over the lifetime of the system, up to the
evolution stage at which the cocoon appeared. In contrast to the
cocoon, evolutionary effects in the spin-down power of the pulsar
become critical in determining the available energy.  The total energy
injected into the ERN depends critically on the assumed birth-period
of the pulsar ($P_0$). The energy in relativistic particles is $\sim
2\times10^{49} \epsilon (P_{0}/30 \mathrm{ms})^{-2}$.
The electron content of the ERN has been estimated to be $5\times
10^{48}$ erg~\citep{fermi:VelaX} using the similarity in radio and GeV
$\gamma$-rays. Birth-periods as high as $P_{0} = 40$ ms have been
discussed~\citep{vanDerSwaluwWu2001} to account for the ratio of PWN
radius to SNR radius but require $\epsilon$ close to 1.  No
explanation has been put forward for the absence of $>100$ GeV particles
in the ERN required by the steep Fermi-LAT spectrum.  Given the
current-day magnetic field of $\sim 4 \mu$G, the cooling times for
particles emitting in the GeV-range are very large and synchrotron
cooling seems implausible. While strong evolution of the magnetic
field in PWNe is possible~\citep[see e.g.][]{Gelfand2009}, an
implausibly high effective time-averaged B-field of $>100 \mu$G within
the whole ERN is required to explain the Fermi-LAT GeV spectrum with a
cooling break.
Since there is no obvious theoretical reason why the maximum
accelerated energy from the pulsar should be time-dependent, the
remaining explanation is that the bulk of the $>100$ GeV particles
have escaped from the ERN into the interstellar medium (ISM). Whilst
confinement of particles in PWNe is thought to be effective in their
early evolution
it is likely that the interaction with the SNR reverse shock which
seems to have appeared in Vela~X a few thousand years
ago~\citep{Blondin:2001p640, Gelfand2009} brings an end to the
effective confinement of electrons. Such a reverse shock interaction
is expected to disrupt the PWN sufficiently, through e.g. the growth
of Rayleigh-Taylor instabilities, that diffusion of particles out of
the PWN becomes possible.
This explanation for the absence of $>100$ GeV particles in the ERN
requires a diffusion coefficient of $D =
R_{\mathrm{pc}}^2/2t_{\mathrm{kyr}} \sim \times 10^{26}$ cm$^2$
s$^{-1}$($R$ in parsec, $t$ in units of kyears).
For typical parameters in Vela~X this is of order 1000 times slower
than that inferred for cosmic rays in the ISM, but much faster than
Bohm diffusion for the magnetic field in the ERN~\citep{felixbook}.

\paragraph{Calculation}
For the calculation of the present-day SED of Vela~X and the energy
and time-dependent flux of escaping particles we employ a Monte-Carlo
approach where propagation and energy losses of individual
particles are considered. Following earlier studies we neglect
adiabatic losses due to the likely return of most of the lost energy
at the time of reverse shock crushing of the PWN~\citep[see
e.g.][]{OkkieVelaX}. Synchrotron and Klein-Nishina (KN) inverse
Compton (IC) losses are calculated at each time step. For the
calculation of IC cooling/radiation we adopt the soft-photon fields
used by~\citet{OkkieVelaX} and~\citet{fermi:VelaX}, with black-body
spectra of temperatures 25~K and 6500~K and energy densities of 0.4
eV cm$^{-3}$. In the first stage of evolution of the system
confinement is assumed to be 100\% effective and a one-zone
approach~\citep[following the treatment in][]{HintonAharonian2007} is
employed. A canonical braking index of 3~\citep{OkkieBraking} is
assumed over the evolution of the pulsar, rather than the value
measured for the recent past~\citep[determined to be $1.6 \pm 0.1$
by][]{Dodson2003}.
The choice of this value has very little impact on the final model
curves. For $P_0 = 30$ms, the implied age of the system is
$t_\mathrm{sys} = 10$ kyears. At $t_\mathrm{rs} = \sim 70\%
t_\mathrm{sys}$~\citep[the age when the reverse shock is thought to
start interaction with the ERN, see e.g.][]{Blondin:2001p640},
diffusive transport is switched on, with particles initially uniformly
distributed in a spherical PWN, with continued injection until
$t_\mathrm{sys} - t_\mathrm{coc}$ ($t_{\mathrm{coc}} = 230$ years,
$\epsilon=0.3$), from which time particles injected at the termination
shock are assumed to be confined within the cocoon. This is consistent
with simulations~\citep[e.g.][]{vanDerSwaluw2004} in that when the
reverse shock has collided with the PWN, the pulsar no longer injects
particles into the ``relic PWN'' but forms a new PWN.  Diffusion is
assumed to be a fixed factor $\delta$ faster than Bohm diffusion, with
the magnetic field in the nebula falling continuously with $t^{-0.5}$
at late times. The free parameters of the model are the pulsar birth
period $P_0$, the efficiency $\epsilon$ of conversion of spin-down
power to particles in the MeV-100 TeV domain, the factor $\delta$ and
the current-day ERN magnetic field. For the TeV nebula the magnetic
field strength and cocoon age form the additional free parameters.  To
model propagation and energy losses of the electrons released into the
ISM we adopt an analytical solution of the diffusion-loss equation
\citep{AtoyanAharonianVoelk1995}, assuming synchrotron and IC
radiation as the dominant loss mechanisms. Synchrotron losses assume a
local ISM magnetic field strength of 5 $\mu G$.  The IC losses were
calculated using the full KN description, assuming 4 radiation fields
(CMB, IR, G-K stars and O-B stars) with a total radiation density of 1
eV cm$^{-3}$.

\section{Discussion}

\begin{figure*}[ht]
\epsscale{1}
\plotone{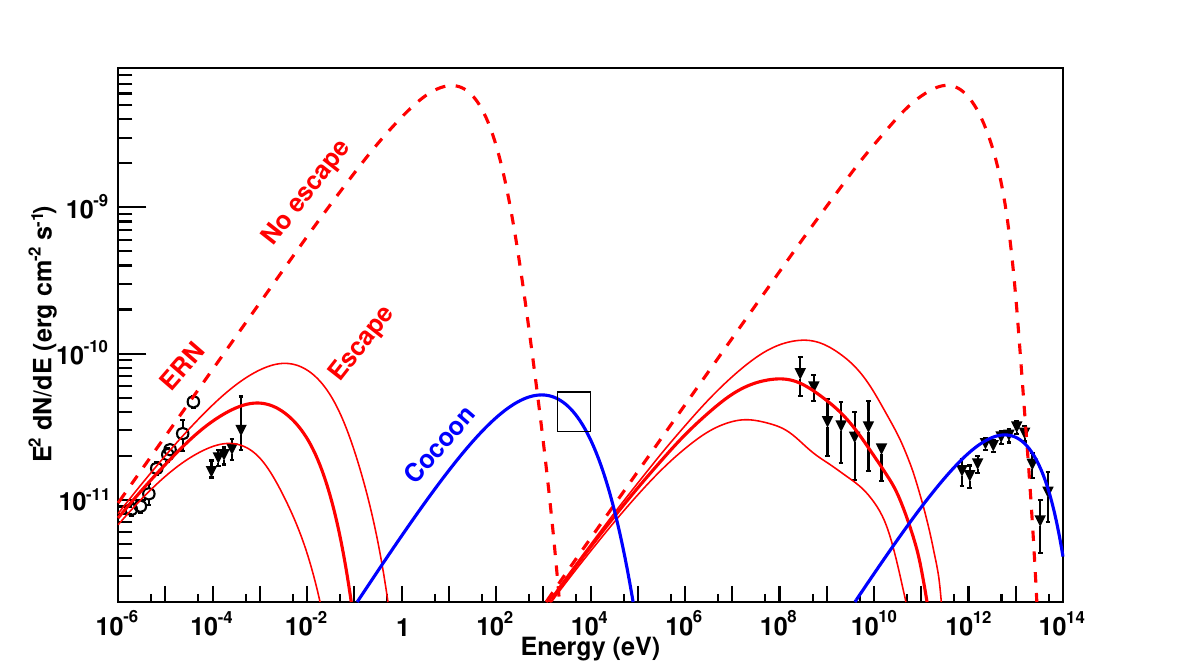}
\caption{ Measured SEDs for the
  cocoon/TeV nebula (blue) and the ERN (red) compared to model
  curves. Radio data are from ~\citet{Alvarez:2001p2311},
  WMAP, ASCA and Fermi-LAT from~\citet{fermi:VelaX} and TeV 
  from~\citet{hess:VelaX}. 
 The total flux  of the TeV nebula is $\sim30$\%
 higher\citep{hess:velaXNew}, but the 
  X-ray flux in this larger region is not well constrained. The
  magnetic field in the cocoon was chosen to be $4.5 \mu$G. Injection
  from the hard X-ray nebula into this region follows a
  power-law with index $-2$ and cutoff at $70$~TeV.  The dashed line
  shows the ERN emission expected in the absence of escape and the
  solid curves for energy-dependent escape with $\delta$=1000, 2000
  (thick line) and 4000 times faster than Bohm diffusion. The
  current-day magnetic field is taken as $4\mu$G. The curves
  assume a birth-period of $P_0 = 22$ ms and 
  efficiency $\epsilon = 0.3$ (almost identical results are
  obtained for $P_0 = 30$ms and $\epsilon = 0.5$) .
\label{fig:2}
}
\end{figure*}

Figure~\ref{fig:2} shows measured and calculated SEDs for the ERN
(red) and the TeV cocoon (blue). There is a striking inconsistency
between the WMAP measurement shown in~\citep{fermi:VelaX} and earlier
radio measurements~\citep{Alvarez:2001p2311,Hales:2004p3641}. We
attempt to approximately match the spectral shapes, rather than the
$\sim 10^{-4}$ eV (50 GHz) normalisation, given the apparent factor
$\sim 3$ disagreement. Data from Planck should resolve this issue in
the near future. The main impact on our model is a change in the
current-day magnetic field.

As previously discussed, cooling effects are not important inside the
cocoon (blue curves). The spectrum in the cocoon is therefore
unmodified from the injection spectrum. The assumption is that the
cooling of the particles inside the cocoon has happened close to the
pulsar, consistent with the low magnetic field ($\sim 4 \mu$G) in the
cocoon and the much higher field ($\sim 100 \mu$G) close to the
pulsar. Thus, this curve is similar to the time-independent
one-zone IC model shown in~\citet{hess:VelaX}.

The expected ERN flux (red) in the absence of particle escape for
constant efficiency $\epsilon$ is shown as dashed lines in
Fig.~\ref{fig:2}, clearly demonstrating how dramatically different the
observed spectrum in the Fermi-LAT energy range is. In this case the
peak energy and fluxes are determined by particle cooling. The solid
lines show the energy-escape based models under the approximation of
spherical symmetry and a 7~pc radius. Different curves illustrate the
effects of different diffusion coefficients. The apparently best-match
to the data yields a factor of $\delta = 2000$ faster than Bohm
diffusion (with the same $D \propto E^{\delta}$, $\delta = 1 $ energy
dependence). Lower exponents $\delta = 0.6$ produce acceptable fits
with modifications of the model parameters within a reasonable range,
whilst significantly lower values seem to be excluded in this
scenario.  We emphasize that these are illustrative models rather than
an attempt at fully describing the system. The details depend
critically on the time- and spatial evolution. The apparent
small-scale differences between the morphology in the radio and GeV
$\gamma$-ray bands (Figure~\ref{fig:1} left) indicate highly
non-uniform magnetic fields in Vela~X.  The true situation of escape
of particles out of the ERN is likely highly complex.

\begin{figure*}[ht]
\epsscale{1.1}
\plottwo{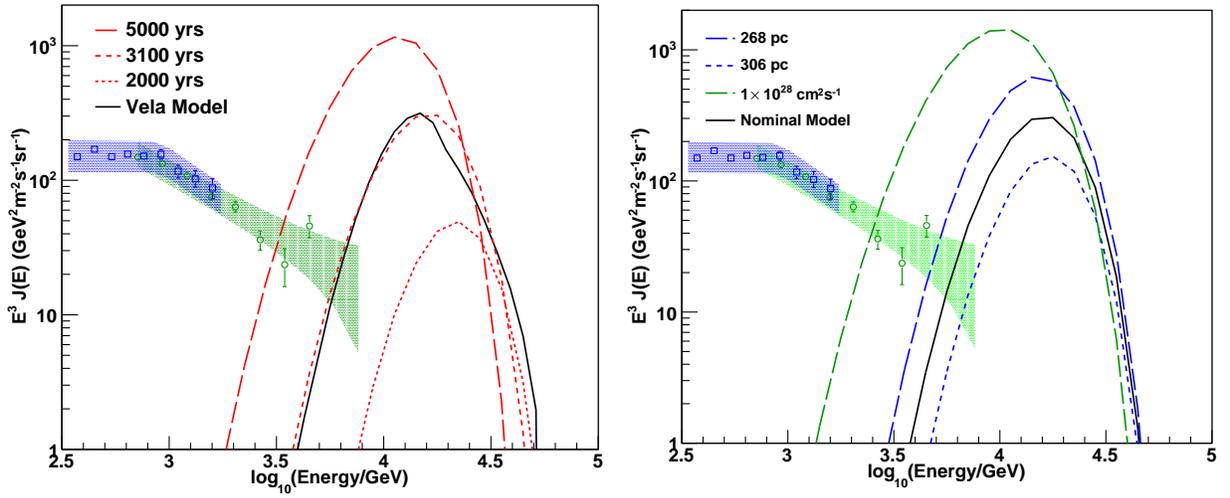}{fig3b.pdf}
\caption{Left: Predicted electron spectrum from Vela~X using the time
  dependent injection model (black). Simple burst-like power-law
  injection spectra are shown for comparison assuming different
  injection times (all d=290 pc, and a diffusion coefficient at 10 GeV
  of $5 \times 10^{27}$ cm$^2$s$^{-1}$ with energy dependence of
  E$^{0.6}$). Right: Effects of varying distance (blue) or diffusion
  coefficient (green) for burst-like injection (at 3100 years).
  Data points show the H.E.S.S.\ electron
  measurements~\citep{hess:electrons,hess:electrons2}. Shaded areas
  represent the systematic uncertainties.
\label{fig:3}
}
\end{figure*}

Figure \ref{fig:3} (left) shows the expected local electron spectrum
from our model, assuming nominal diffusion parameters (black line, see
caption for details). Several key parameters affect the local CR
electron signature of a source: the distance, the electron injection
time and the diffusion coefficient. We investigate the uncertainty on
these parameteres by using a simple burst-like injection into the ISM
and a spectrum adapted to match the signature of the full calculation
($E^{-1.8} \exp\left( E /5\, {\mathrm{TeV}}\right)$ with an energy
input of $6.8 \times 10^{48}$ erg). Figure~\ref{fig:3} (left) shows in
red the effect of varying the time at which this burst of electrons is
released into the ISM. It can be seen that for an injection time
$t_{\mathrm{inj}} = 3$ ky, burst-like injection
produces similar results to the time-dependent injection model.
Figure~\ref{fig:3} (right) shows the effects of varying the distance
to Vela~X within the allowed errors (blue curves)~\citep{Dodson2003}
and varying the diffusion coefficient (green curve). If source and
diffusion parameters are changed from the nominal model it becomes
difficult to keep the spectrum consistent with the H.E.S.S.\
measurements.
Due to the very recent injection of a large number of high energy
electrons, Vela~X will produce a distinct feature in the local
electron spectrum.  Such a signature should be easily observable by
instruments like CTA~\citep{CTA}, expected to achieve a factor $>10$
increase compared to H.E.S.S. in collection area for
well-contained/high-telescope-multiplicity events. Combined with an
effective FoV solid angle increase of a factor $>4$ and $\sim 5$ times
longer observation time (the published H.E.S.S.\ data represents $\sim
150$h), the expected increase in electron statistics (with comparable
systematics and background levels) is a factor $\sim 300$, extending
the spectrum from $\sim 4$TeV up to $\sim 70$TeV for an E$^{-3}$
spectrum.

\section{Conclusions}

Due to its proximity and well-known distance, Vela~X presents a unique
opportunity to constrain models using excellent spectral and
morphological measurements and the locally-measured electron/positron
flux. The model presented here, of energy-dependent escape from
Vela~X, produces a clear signature in the CR electron spectrum, which
could be measured in detail with CTA. If this scenario proves
correct and Vela~X proves to be typical of PWN of this age, the
release of $>$ TeV particles from PWN relatively early in their
evolution may help to explain the rise in the CR positron fraction
seen at high energies and at the same time help understand the absence
of the significant population of Fermi "relic"-PWN that had been
predicted before launch \citep[see e.g.][]{okkieBirthPeriod}.

\acknowledgements{

  We thank A.~Hales for providing the 8.4 GHz reprocessed image of
  (originally provided by D.~Bock). We thank Felix Aharonian and an
  anonymous referee for helpful comments. SO acknowledges
  support by a Humboldt foundation Feodor-Lynen fellowship.  }


\begin{thebibliography}{43}

\bibitem[{Abdo} {et~al.}(2010)]{fermi:VelaX}
{Abdo}, A.~A., {Ackermann}, M., {Ajello}, et al., 2010, \apj,
  713, 146

\bibitem[{Abdo} {et~al.}(2009)]{fermi:electrons}
{Abdo}, A.~A., {Ackermann}, M., {Ajello}, M., et al., 2009, PRL, 102, 181101

\bibitem[{Ackermann} {et~al.}(2011)]{fermi:positrons}
{Ackermann}, M., {Ajello}, M., {Allafort}, A., et al., 2011, ArXiv/1109.0521

\bibitem[{Adriani} {et~al.}(2009)]{pamela:positrons}
{Adriani}, O., {Barbarino}, G.~C., {Bazilevskaya}, G.~A., et al., 2009, Nature, 458, 607

\bibitem[{{Aharonian} {et~al.}(1997){Aharonian, F.~A., Atoyan,
      A.~M. \& Kifune, T.}}]{AharonianAtoyanKifune1997}
{Aharonian}, F.~A., {Atoyan}, A.~M.,  \& {Kifune}, T., 1997, \mnras, 291, 162

\bibitem[{{Aharonian}(2004)}]{felixbook}
{Aharonian}, F.~A. 2004, {Very High Energy Cosmic Gamma Radiation} ({World
  Scientific Pub Co Inc})

\bibitem[{Aharonian} {et~al.}(2008)]{hess:electrons}
{Aharonian}, F., {Akhperjanian}, A.~G., {Barres de Almeida}, U., et al., 2008,
  PRL, 101, 261104

\bibitem[{Aharonian} {et~al.}(2006)]{hess:VelaX}
{Aharonian}, F., {Akhperjanian}, A.~G., {Bazer-Bachi}, A.~R., et al., 2006, \aap,
  448, L43

\bibitem[{Alvarez {et~al.}(2001)Alvarez, Aparici, May, \&
  Reich}]{Alvarez:2001p2311}
Alvarez, H., Aparici, J., May, J., \& Reich, P. 2001, A{\&}A, 372, 636

\bibitem[{Aschenbach {et~al.}(1995)Aschenbach, Egger, \&
  Trumper}]{Aschenbach:1995p669}
Aschenbach, B., Egger, R., \& Trumper, J. 1995, \nat, 373, 587

\bibitem[{{Atoyan} {et~al.}(1995){Atoyan}, {Aharonian}, \&
  {V{\"o}lk}}]{AtoyanAharonianVoelk1995}
{Atoyan}, A.~M., {Aharonian}, F.~A., \& {V{\"o}lk}, H.~J. 1995, \prd, 52, 3265

\bibitem[{Blondin {et~al.}(2001)Blondin, Chevalier, \&
  Frierson}]{Blondin:2001p640}
Blondin, J.~M., Chevalier, R.~A., \& Frierson, D.~M. 2001, \apj, 563, 806

\bibitem[{{B{\"u}sching} {et~al.}(2008){B{\"u}sching}, {de Jager}, {Potgieter},
  \& {Venter}}]{Buesching2008}
{B{\"u}sching}, I., {de Jager}, O.~C., {Potgieter}, M.~S., \& {Venter}, C.
  2008, \apjl, 678, L39

\bibitem[{{Caraveo} {et~al.}(2001){Caraveo}, {De Luca}, {Mignani}, \&
  {Bignami}}]{Caraveo:2001p930}
{Caraveo}, P.~A., {De Luca}, A., {Mignani}, R.~P., \& {Bignami}, G.~F. 2001,
  \apj, 561, 930

\bibitem[{{Chi} {et~al.}(1996){Chi}, {Cheng}, \& {Young}}]{ChiChen1996}
{Chi}, X., {Cheng}, K.~S., \& {Young}, E.~C.~M. 1996, \apjl, 459, L83+

\bibitem[{{de Jager}(2007)}]{OkkieBraking}
{de Jager}, O.~C. 2007, \apj, 658, 1177

\bibitem[{{de Jager}(2008)}]{okkieBirthPeriod}
---. 2008, \apjl, 678, L113

%
\bibitem[{{de Jager} {et~al.}(2008){de Jager}, {Slane}, \&
  {LaMassa}}]{OkkieVelaX}
{de Jager}, O.~C., {Slane}, P.~O., \& {LaMassa}, S. 2008, \apjl, 689, L125

\bibitem[{Dodson {et~al.}(2003)Dodson, Legge, Reynolds, \&
  McCulloch}]{Dodson2003}
Dodson, R., Legge, D., Reynolds, J.~E., \& McCulloch, P.~M. 2003, The
  Astrophysical Journal, 596, 1137

\bibitem[{Dubner {et~al.}(1998)Dubner, Green, Goss, Bock, \&
  Giacani}]{Dubner:1998p3619}
Dubner, G.~M., Green, A.~J., Goss, W.~M., Bock, D. C.-J., \& Giacani, E. 1998,
  \aj, 116, 813

\bibitem[{Duncan {et~al.}(1996)Duncan, Stewart, Haynes, \&
  Jones}]{Duncan:1996p3426}
Duncan, A.~R., Stewart, R.~T., Haynes, R.~F., \& Jones, K.~L. 1996, \mnras,
  280, 252

\bibitem[{Dwarakanath(1991)}]{Dwarakanath:1991p3369}
Dwarakanath, K.~S. 1991, \aap, 12, 199

\bibitem[{{Egberts} et al.(2011)}]{hess:electrons2}
{Egberts}, K. et al. (H.E.S.S.~Collaboration), 2011, NIM A, 630, 36

\bibitem[{Frail {et~al.}(1997)Frail, Bietenholz, Markwardt, \&
  Oegelman}]{Frail:1997p3296}
Frail, D.~A., Bietenholz, M.~F., Markwardt, C.~B., \& Oegelman, H. 1997, \apj,
  475, 224

\bibitem[{{Gelfand} {et~al.}(2009){Gelfand}, {Slane}, \& {Zhang}}]{Gelfand2009}
{Gelfand}, J.~D., {Slane}, P.~O., \& {Zhang}, W. 2009, \apj, 703, 2051

\bibitem[Hales {et~al.}(2004)]{Hales:2004p3641}
Hales, A.~S., Casassus, S., Alvarez, H., et al., 2004, \apj, 613, 977

\bibitem[{{Hermann} {et~al.}(2007){Hermann}, {Hofmann}, {Schweizer}, {Teshima},
  \& {CTA consortium}}]{CTA}
{Hermann}, G., {Hofmann}, W., {Schweizer}, T., {Teshima}, M., \& {CTA
  consortium}, f.~t. 2007, ArXiv e-prints, 709

\bibitem[{Hinton \& Aharonian(2007)}]{HintonAharonian2007}
Hinton, J.~A. \& Aharonian, F.~A. 2007, \apj, 657, 302

\bibitem[Dubois {et~al.}(2009)]{hess:velaXNew}
Dubois, F., B. Gl{\"u}ck, Jager, O. C.~D., et al., 2009, Proceedings
of the 31st ICRC in Lodz, 45 

\bibitem[{Kahn {et~al.}(1985)Kahn, Gorenstein, Harnden, \&
  Seward}]{Kahn:1985p3493}
Kahn, S.~M., Gorenstein, P., Harnden, F.~R., \& Seward, F.~D. 1985, \apj, 299,
  821

\bibitem[{{Kobayashi} {et~al.}(2004){Kobayashi}, {Komori}, {Yoshida}, \&
  {Nishimura}}]{Kobayashi2004}
{Kobayashi}, T., {Komori}, Y., {Yoshida}, K., \& {Nishimura}, J. 2004, \apj,
  601, 340

\bibitem[{Large {et~al.}(1968)Large, Vaughan, \& Mills}]{Large:1968p3223}
Large, M.~I., Vaughan, A.~E., \& Mills, B.~Y. 1968, \nat, 220, 340

\bibitem[{{Manchester} {et~al.}(2005){Manchester}, {Hobbs}, {Teoh}, \&
  {Hobbs}}]{atnf}
{Manchester}, R.~N., {Hobbs}, G.~B., {Teoh}, A., \& {Hobbs}, M. 2005, \aj, 129,
  1993

\bibitem[{Mangano {et~al.}(2005)Mangano, Massaro, Bocchino, Mineo, \&
  Cusumano}]{Mangano:2005p2719}
Mangano, V., Massaro, E., Bocchino, F., Mineo, T., \& Cusumano, G. 2005, \aap,
  436, 917

\bibitem[{Markwardt \& \"Ogelman(1995)}]{Markwardt:1995p3614}
Markwardt, C.~B. \& \"Ogelman, H. 1995, \nat, 375, 40

\bibitem[{Markwardt \& \"Ogelman(1997)}]{Markwardt:1997p3699}
Markwardt, C.~B. \& \"Ogelman, H.~B. 1997, \apjl, 480, L13

\bibitem[{Moriguchi {et~al.}(2001)Moriguchi, Yamaguchi, Onishi, Mizuno, \&
  Fukui}]{Moriguchi:2001p3627}
Moriguchi, Y., Yamaguchi, N., Onishi, T., Mizuno, A., \& Fukui, Y. 2001, \pasj,
  53, 1025

\bibitem[{Nishimura} {et~al.}(1980)]{Nishimura1980}
{Nishimura}, J., {Fujii}, M., {Taira}, T., et al., 1980, \apj, 238, 394

\bibitem[{{Profumo}(2008)}]{Profumo2008}
{Profumo}, S. 2008, ArXiv e-prints

\bibitem[{{Shen}(1970)}]{Shen1970}
{Shen}, C.~S. 1970, \apjl, 162, L181+

\bibitem[{{van der Swaluw} {et~al.}(2004){van der Swaluw}, {Downes}, \&
  {Keegan}}]{vanDerSwaluw2004}
{van der Swaluw}, E., {Downes}, T.~P., \& {Keegan}, R. 2004, \aap, 420, 937

\bibitem[{{van der Swaluw} \& {Wu}(2001)}]{vanDerSwaluwWu2001}
{van der Swaluw}, E. \& {Wu}, Y. 2001, \apjl, 555, L49+

\bibitem[{Weiler \& Panagia(1980)}]{Weiler:1980p1759}
Weiler, K.~W. \& Panagia, N. 1980, A{\&}A, 90, 269

\bibitem[{{Y{\"u}ksel} {et~al.}(2009){Y{\"u}ksel}, {Kistler}, \&
  {Stanev}}]{YukselKistler2009}
{Y{\"u}ksel}, H., {Kistler}, M.~D., \& {Stanev}, T. 2009, Physical Review
  Letters, 103, 051101

\end{thebibliography}

\end{document}